\documentclass[twocolumn,runningheads]{svjour2}
\smartqed  
\usepackage{graphicx}
\usepackage{mathptmx}   
\usepackage{amsmath}
\usepackage{amsfonts}
\usepackage{amssymb}
\usepackage{latexsym}
\usepackage[authoryear]{natbib}

\usepackage{journals}

\journalname{Astrophysics and Space Science (CoRoT/ESTA Volume)}
\newcommand{\msol}{\mbox{${M}_{\odot}$}}

\newcommand{\corot}{{\small CoRoT}}

\newcommand{\esta}{{\small ESTA}}

\newcommand{\CESAM}{{\small\bf CESAM}}
\newcommand{\CLES}{{\small CLES}}

\newcommand{\ADIPLS}{{\small ADIPLS}}
\newcommand{\LOSC}{{\small LOSC}}
\newcommand{\POSC}{{\small POSC}}

\newcommand{\cesam}{{\small CESAM}}
\newcommand{\cesamk}{{\small CESAM2k}}

\newcommand{\ZAMS}{{\small ZAMS}}

\newcommand{\adipls}{{\small ADIPLS}}
\newcommand{\thisapss}{Astrophys. Space Sci. (CoRoT/ESTA Volume)}

\newcommand{\etal}{{\em et al.}}

\begin{document}

\title{Reference grids of stellar models and oscillation frequencies: 
}
\subtitle{Data from the CESAM stellar evolution code and ADIPLS oscillation programme}

\titlerunning{CESAM-ADIPLS reference stellar models and oscillation frequencies}        

\author{Yveline Lebreton        \and
        Eric Michel 
}


\institute{Y. Lebreton \at
               Observatoire de Paris, GEPI, CNRS UMR 8111, 5 Place Janssen, 92195 Meudon, France \\
              Tel.: +33-299236386\\
              Fax: +33-299236957\\
              \email{Yveline.Lebreton@obspm.fr}           
           \and
           E. Michel \at
              Observatoire de Paris, LESIA, CNRS UMR 8109, 5 Place Janssen, 92195 Meudon, France
}

\date{Received: date / Accepted: date}

\maketitle
\begin{abstract}
We present grids of stellar models and their associated oscillation frequencies that have been used by the \corot\ \textit{Seismology Working Group} during the scientific pre\-paration of the \corot\ mission. The stellar models have been calculated with the \cesam\ stellar internal structure and evolution code while the oscillation frequencies have been obtained from the \cesam\ models by means of the \adipls\ adiabatic oscillation programme. The grids cover a  range of masses, chemical compositions and evolutionary stages corresponding to those of the \corot\ primary targets. The stellar models and oscillation frequencies are available on line through the \textit{Evolution and Seismic Tools Activity} (ESTA) web site.


\end{abstract}

\section{Introduction}\label{intro}

During the scientific preparation of the \corot\ mission, a reference frame in the H--R diagram had to be provided for the \corot\ potential target stars. In particular, the critical selection of the primary targets and the census and characterisation of the stellar content in the fields surrounding the potential primary targets have required to locate all considered stars (or stellar systems) in the H--R diagram and eventually to estimate their mass, evolutionary stage and expected oscillation spectra. For that purpose we have calculated the internal structure and evolution of stars in a range of masses, chemical compositions and evolution stages of interest for the modelling of \corot\ targets \citep{2005sf2a.conf..283M}. We have used the {\cesam}\footnote{Code d'Evolution Stellaire Adaptatif et Modulaire} stellar evolution code \citep{Morel97,pm-apss}. We have selected several models along each evolution sequence and we have calculated their oscillation frequencies by means of the \adipls\footnote{Aarhus Adiabatic Pulsation Package), available at \tt\small http://astro.phys.au.dk/jcd/adipack.n} programmes \citep{1982MNRAS.199..735C, jcd2-apss}.

In this paper we briefly present these grids of models and oscillation frequencies. Section~\ref{sec:models} presents the {\cesam} numerical code and the input physics and initial parameters used to calculate the grids while Section~\ref{sec:osc} presents the oscillation data obtained from the {\adipls} adiabatic oscillation code. The present models and associated oscillation frequencies have been made available on the ESTA web site\footnote{\tt\small http://www.astro.up.pt/corot/models/ and also on the Paris Observatory web site at \tt\small http://wwwusr.obspm.fr/$\sim$lebreton/Modeles/CESAM\_COROT.html}.

\section{CoRoT/CESAM stellar models}\label{sec:models}

\subsection{Input physics and numerical aspects}

The {\cesam} stellar evolution code has been written and developed by P. Morel and collaborators since the late 1980's. The general description of the code is given by \citet{pm-apss} but see also \citet{Morel97}. {\cesam} is a public code. The source is available on-line at the web site -- {\tt http://www.obs-nice.fr/cesam} -- together with a short guide of directions for use and a comprehensive document giving a complete description of the numerics and of the physics implemented. All the models presented here were calculated {in 2003} with the 4.2 version of {\cesam} written in Fortran77. We point out that the present version of {\cesam}, {\cesamk}, was re-programmed in Fortran95; it has also been updated with new input physics and has been enriched with the inclusion of new physical processes \citep{pm-apss}.

The models have be initialised on the zero age main sequence ({\ZAMS}) and the evolution has been followed up to the beginning of the red giant branch. The models have typically 600 mesh points in the interior while the atmosphere is restored on a grid of typically 100 grid points. During hydrogen burning the time step is adjusted according to the relative changes of the hydrogen abundances. In the present calculations it takes from 165 to 420 time steps to evolve from the {\ZAMS} to the ascent of the red giant branch, depending on the mass.

All the models have been calculated with the same given set of standard input physics. Here, we have neglected microscopic and turbulent diffusion processes as well as rotation or magnetic fields. The input physics are the following:

\begin{itemize}
\item \textit{Equation of State.} We used the so-called {CEFF} analytical equation of state \citep{ceff92} which is an improved version of the {EFF} \citep{eff73} equation of state including the Coulomb corrections to pressure in the so-called Debye-H\"uckel approximation.

\item \textit{Opacities.} We used the OPAL95 opacity tables \citep{ir96} complemented at low temperatures ($T\lessapprox 10^4 K$) by the \citet{af94} tables. The metal mixture of the opacity tables is fixed to the initial mixture of the model (see below).

\item \textit{Nuclear reaction rates.}  We used the basic pp and CNO reaction networks up to the $^{17}O(p,\alpha)^{14}N$ reaction. In the present models the \cesam\ code takes $^7Li$, $^7Be$ and $^2H$ at equilibrium. The nuclear reaction rates are from \cite{CF88}. Weak screening is assumed under Salpeter's formulation \citeyearpar{1954AuJPh...7..373S}.

\item \textit{Convection and overshooting.} We use the classical mixing length treatment of  \citet{1958ZA.....46..108B} under the formulation of \citet{1965ApJ...142..841H} taking into account the optical thickness of the convective bubble. To find the location of the onset of convection, we used  Schwarz\-schild's criterion. In models calculated with overshooting, the convective core is extended on a distance $l_{\rm ov}=\alpha_{\rm ov}\times \min(H_{\rm p}, R_{\rm cc})$ where $\alpha_{\rm ov}$ is the overshooting parameter, $H_{\rm p}$ the pressure scale height and $R_{\rm cc}$ the radius of the convective core. The core is mixed in the region corresponding to both the convective and overshooting regions and in the overshooting region the temperature gradient is taken to be equal to the adiabatic gradient. 

\item \textit{Atmosphere.} Eddington's grey $T(\tau)$ law is used for the atmosphere calculation. The hydrostatic equation is integrated in the atmosphere, starting at the optical depth $\tau=10^{-4}$ and the connection with the envelope is made at $\tau=10$ to ensure the validity of the diffusion approximation
\citep{Morel94}. At this level, we insure the continuity of the variables and of their first derivatives with space.  The radius of the star is taken to be the bolometric radius, i.e. the radius at the level where the local temperature equals the effective temperature ($\tau=2/3$ for the Eddington's law).

\item \textit{Initial abundances of the elements and heavy elements mixture.} All models were calculated with the classical GN93 solar mixture of heavy elements from \citet{GN93} corresponding to a solar metals to hydrogen ratio $(Z/X)_\odot=0.0245$ where $Z$ and $X$ are the abundances in mass fraction of heavy elements and hydrogen respectively. In the nuclear reaction network the initial abundance of each chemical species is split between its isotopes according to the isotopic ratio of nuclides. The initial amount of $^2H$ is assumed to be converted in $^3He$.
\end{itemize}

\begin{figure}[ht!]
\centering
\resizebox*{0.9\hsize}{!}{\includegraphics*{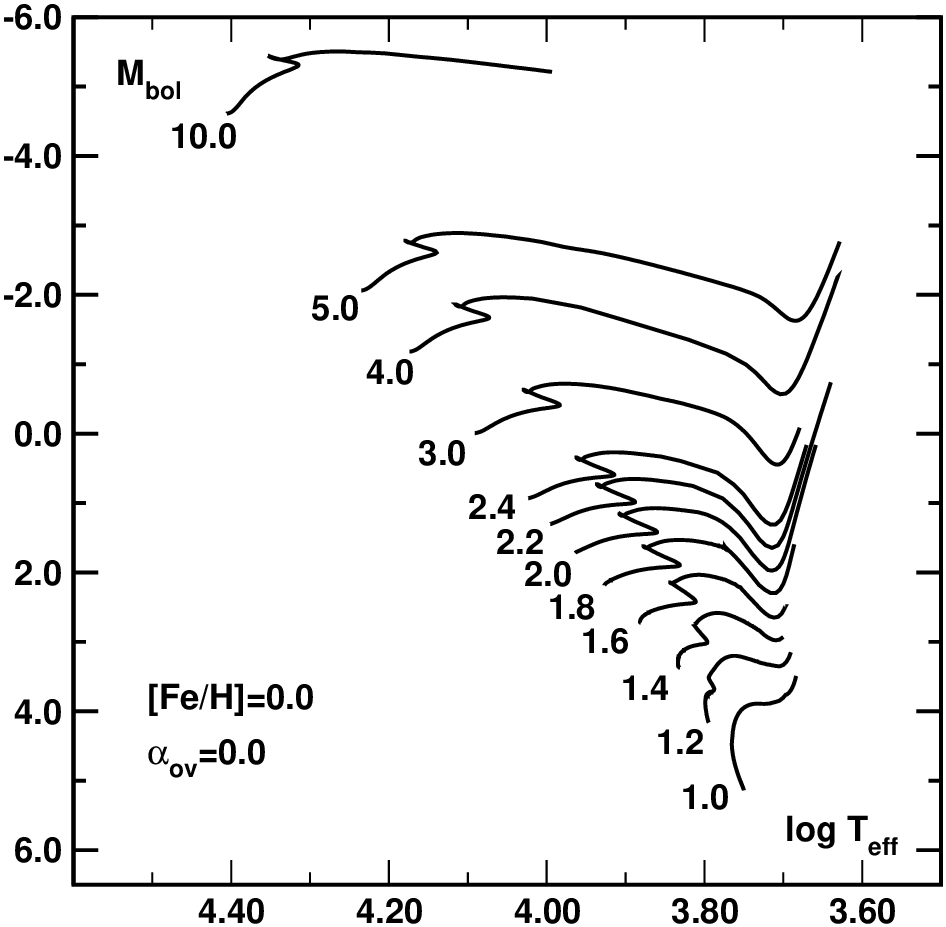}}
\caption{Grid {\bf G1}: evolutionary sequences in the H--R diagram for models with solar metallicity $\rm [Fe/H]=0.0$ corresponding to $Z/X=0.0245$, helium abundance $Y=0.2672$, no overshooting.}
\label{fig:feh00}       
\end{figure}

\subsection{Initial parameters of the models}

The calibration of a solar model in luminosity and radius with the input physics given above leads to an initial abundance of helium $Y_\odot=0.2672$, and to a solar mixing-length parameter $\alpha_{\rm MLT, \odot}=1.63$. The corresponding initial hydrogen  and metal abundances are respectively $X_\odot=0.7153$ and $Z_\odot=0.0175$.

We have calculated models for two values of the $\rm [Fe/H]$ ratio: $\rm [Fe/H]=0.0$ (solar) and $-0.1$ (metal deficient). The metallicity of the models $Z$ has been obtained from the relation ${\rm [Fe/H]}=\log(Z/X)-\log(Z/X)_\odot$. The individual abundances of metals have been scaled on the solar ones. 

For models with non solar metallicity, we have derived the helium abundance in mass fraction $Y$ according to the relation $Y=Y_{\rm p}+Z(\Delta Y/\Delta Z)$, where $\Delta Y/\Delta Z$ is the helium to metal enrichment ratio and $Y_{\rm p}$ is the primordial helium abundance. We took $\Delta Y/\Delta Z=(Y_{\odot}-Y_{\rm p})/Z_{\odot}\sim 1.3$ and $Y_{\rm p}=0.245$ \citep[see][]{2004ApJ...617...29O}. 

All the models were calculated with a mixing-length parameter equal to the solar one. Finally, for solar metallicity we calculated two grids of models, one without overshooting and one including overshooting with the value $\alpha_{\rm ov}=0.15$.

We have calculated models with masses in the range covered by the {\corot} targets: $1.0$, $1.2$, $1.4$, $1.6$, $1.8$, $2.0$, $2.2$, $2.4$, $3.0$, $4.0$, $5.0$ and $10.0$ {\msol}.

\begin{figure}[hb!]
\centering
\resizebox*{0.9\hsize}{!}{\includegraphics*{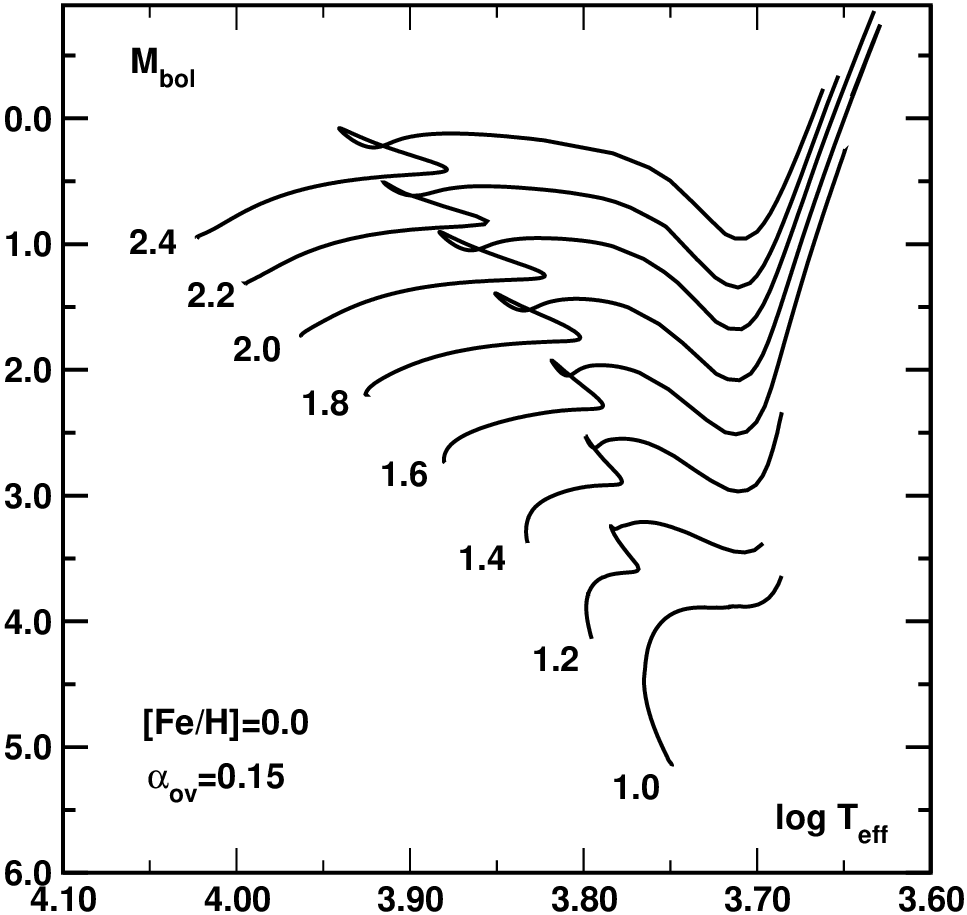}}
\caption{Grid {\bf G2}: evolutionary sequences in the H--R diagram for models with solar metallicity $\rm [Fe/H]=0.0$ corresponding to $Z/X=0.0245$, helium abundance $Y=0.2672$ and overshooting $\alpha_{\rm ov}=0.15$}
\label{fig:feh00ov}       
\end{figure}

\subsection{Model grids}

\begin{figure}[ht!]
\centering
\resizebox*{0.9\hsize}{!}{\includegraphics*{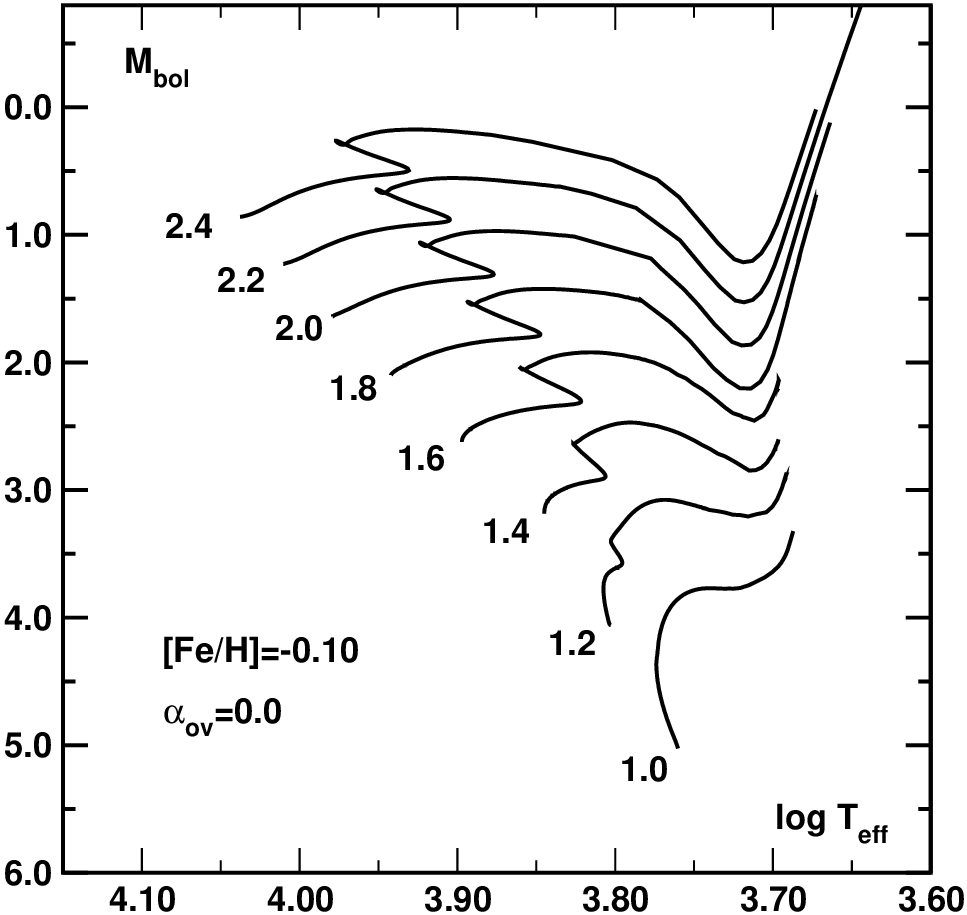}}
\caption{Grid {\bf G3}: evolutionary sequences in the H--R diagram for models with sub-solar metallicity $\rm [Fe/H]=-0.1$ corresponding to $Z/X=0.0194$, helium abundance $Y=0.2632$ and no overshooting}
\label{fig:fehm01}       
\end{figure}

We present three grids of models. Grid {\bf G1} consists in evolutionary sequences for the 12 values of the stellar mass listed above (between $1.0$ and $10.0$ {\msol}), a solar metallicity and no overshooting. { This grid covers the domain of mass of main {\corot} targets, including solar-like oscillators and $\delta$ Scuti and $\beta$ Cephei pulsators. In addition to this standard grid, we have calculated two grids restricted to solar-type and $\delta$ Scuti stars, i.e. for the 8 lower values of masses listed above, between $1.0$ and $2.4$ {\msol}. Grid {\bf G2} consists in  evolutionary sequences  with solar metallicity and overshooting  ($\alpha_{\rm ov}=0.15$) and grid {\bf G3} consists in  evolutionary sequences with $[\rm Fe/H]=-0.1$ and no overshooting}. Figures~\ref{fig:feh00}, \ref{fig:feh00ov} and \ref{fig:fehm01} display the evolutionary sequences in the H--R diagram for the 3 grids of {\cesam} models available on the {\esta} web site\footnote{\tt\small http://www.astro.up.pt/corot/models/}. For each evolutionary sequence, we also provide on the web site a file in which the properties of the models are displayed: mass, logarithm of the effective temperature $\log T_{\rm eff}$, bolometric magnitude, logarithm of the luminosity $\log(L/L_\odot)$, logarithm of the surface gravity $log\ {\rm g}$ and age.

\section{ADIPLS oscillation frequencies}\label{sec:osc}

Using the {\adipls} stellar oscillation program, we have calculated the oscillation frequencies for a selection of models of the {\bf G2} grid (i.e. the grid corresponding to models with solar metallicity $\rm [Fe/H]=0.0$, $Z/X=0.0245$, helium abundance $Y=0.2672$ and overshooting $\alpha_{\rm ov}=0.15$). In relation with {\corot} targets, we have chosen two values of the mass in the {\bf G2} grid:  $M=1.2$ {\msol} which is { representative of solar like oscillators} and $M=1.8$ {\msol} which is { representative of $\delta$ Scuti pulsators}. For the $M=1.2$ {\msol} sequence, we have selected models along the evolutionary track by steps of $100\ \rm Myr$, except on the hook at the end of the main-sequence  where steps are of $10\ \rm Myr$. For the $M=1.8$ {\msol} sequence, we have selected models by steps of $50\ \rm Myr$ on the main--sequence and by steps of $10\ \rm Myr$ from the red point to the bottom of the red giant branch.

For each mass we provide two archives on the {\esta} web site, a {\bf Oi} and a {\bf Fi} archive, where i=1 and 2 for $M=1.2$ and  $M=1.8$ {\msol} respectively.

The {\bf Oi} archive contains ASCII files, with a {\em{.osc}} extension, each corresponding to a selected model on the evolutionary sequence. The {\em{.osc}} files can be read by means of the programme tools available in the {\adipls} package and described on the {\esta} web site\footnote{at {\tt\small http://www.astro.up.pt/corot/ntools/modconv}}. They list the global properties of the considered model (name, mass, age, initial composition, luminosity, photospheric radius, etc.) as well as a wide range of model variables at each mesh point (distance r to the centre, mass inside
the sphere of radius r, pressure, density, temperature, chemical composition, opacity and several other physical
parameters including quantities of interest for the computation of adiabatic oscillations like the Brunt-V\"ais\"al\"a frequency).

The {\bf Fi} archive contains ASCII files, with a {\em{.freq}} extension, each corresponding to a selected model on the evolutionary sequence. The {\em{.freq}} files are the result of the frequency calculation by the {\adipls} program. They list the mass, radius central pressure and density of the model and the frequency of oscillation, for each mode identified by its degree $\ell$ and order {\it n}. We point out that these files are not corrected for problems of numbering of the modes that occur for modes of $\ell=1$ in the region of the fundamental. 

\section{Conclusion}\label{sec:conclusion}

{ On the {\esta} web site we have provided grids of stellar models and oscillation frequencies that have been used 
by the \corot\ \textit{Seismology Working Group} during the scientific preparation of the \corot\ mission. The stellar models are provided for a range of masses between $1.0$ and $10.0$ {\msol} and chemical compositions [Fe/H]=$0.0$ and $-1.0$ dex representative of the main  {\corot} targets (solar-type stars, $\delta$ Scuti stars and $\beta$ Cephei stars). Oscillation frequencies are provided all along the main-sequence track for a solar-type oscillator ($M=1.2$ {\msol}) and a $\delta$ Scuti pulsator ($M=1.8$ {\msol}). Up to now these reference grids have been used to locate {\corot} potential targets in the H--R diagram in the process of target selection \citep[see Fig. 1 in][]{yl1-apss} and to study some $\delta$ Scuti candidates for {\corot} \citep{2005AJ....129.2461P}.

On the {\esta} web site, other grids of stellar models and oscillation frequencies can be found. These stellar models have been calculated either with the {\CESAM} or the {\CLES} code and the oscillation frequencies have been calculated either with the {\ADIPLS}, {\POSC} or {\LOSC} code. All this material (codes and grids)  is described in this volume by \citet{yl1-apss, jm1-apss,mmf-apss}. We point out that the thorough comparisons of the stellar evolutionary codes and oscillations codes performed under {\esta} have shown very good agreement between models calculated by the {\CESAM} and {\CLES} codes which gives us confidence in the use of the present grids \citep[see][]{yl3-apss,jm2-apss,moya-apss}.
}

\end{document}